\documentclass[amssymb,12pt]{article}
\usepackage{graphicx}
\usepackage{amsmath}
\usepackage{amsfonts}
\usepackage{amssymb}

\begin{document}
\title{{\bf {Analysis of the Drift Instability Growth Rates in Non-ideal
Inhomogeneous Dusty Plasmas }}}
\author{C. Cereceda, \underline{J. Puerta}, P. Mart\'{\i}n and E. Castro}
\maketitle

\begin{abstract}
\indent In this paper we introduce an algebraic form of the
dispersion relation for a non ideal inhomogeneous dusty plasma in
order to improve drastically the calculation of the drift
instability growth rate. This method makes use of the multipole
approximation of the Z dispersion function, previously published,
and valid for the entire range. A careful analysis of the
solutions spectra of this kind of polynomial equation permits us
to calculate easily the growth rate of the drift instability for
the ion-dust and dust acoustic mode. The value of the parallel to
magnetic field wavelength for which the instability reaches the
maximal value is carefully localized and discussed. The unstable
dust-ion and dust acoustic mode are discriminated and analyzed in
function of the density gradient, Te/Ti - ratio, and dust grain
radius.
\end{abstract}
\maketitle{\centerline{\ Departamento de F\'{\i}sica, Universidad
Sim\'on Bol\'{\i}var, Apdo. 89000,}} {\centerline {Caracas,
Venezuela.}} {\centerline {\ E-mail: cereceda@usb.ve,
jpuerta@usb.ve, pmartin@usb.ve and ecastro@usb.ve }} \indent
\newline

\section{Introduction}

Plasma inhomogeneities across the magnetic field in the presence
of finite - size charged grains causes a wide class of
instabilities of an inhomogeneous dusty plasma called gradient
instabilities. Such instabilities can be studied in the
approximation on magnetic field where we have parallel straight
field lines in order to simplify our treatment. We look for
instabilities in the very low frequency regime where a new
spectrum instabilities and waves appear, induced by the dust
collective dynamics: Dust - Acoustic - Waves (DAWs), Dust - Ion -
Acoustic - Waves (DIAWs), etc. The frequency of DAWs are around 10
Hz as determined in the laboratory and lower in astrophysical
plasmas [1,2]. In the case that grains are in the micron range we
expect a non - ideal behavior due to the fact that the particulate
are highly charged and intermolecular forces could play certainly
an important role. In order to discuss this problem we compare the
ideal properties with the simple hard - core model and in a next
work we will use a better model by means of of the square - well
model and the Pad\'{e} rational approximant to the equation of
state [3] for hard - sphere gas, that in our knowledge is more
realistic as the simple application of the Van der Waals equation
of state [4]. In this paper we show an analysis of the
electrostatic waves and instabilities growth rates in a weakly non
- ideal magnetized dusty plasma with density and temperature
gradients, ignoring charge fluctuation. As introduced before, the
non - ideal behavior is characterized by the hardcore model
defined by
\[
p=n T(1+b_{o}n),
\]
or in similar manner by the square - well model given by the Ree
and Hoover expression [5 ].

\section{Theoretical Model}

In this paper we introduce a new numerical treatment in
combination with a more realistic formulation of the equation of
state to simulate weak non ideal effects in order to analyze
inhomogeneous Vlasov - Dusty Plasma systems where a linearized
dispersion relation is obtained. Due to the lower frequency range
($\omega ,k_{z}v_{T}\ll \omega _{c}$), enough energy can be
transferred from the particle to the wave and instabilities can be
generated. In order to get an adequate linear dispersion relation
with a magnetic field given by ${\mathbf{B}}=B_{0}\hat{\bf{k}}$
for Maxwellian multi-species plasmas (electron, ion and dust), we
introduce our well known and very accurate multipolar
approximation [6] for the ${\bf Z}$ dispersion function.
\newline
In the presence of a magnetic field we have the distribution function of the
species $\alpha $, solution for the kinetic equation

\begin{equation}
\frac{df_{\alpha }}{dt}=\frac{q_{\alpha }}{m_{\alpha }}{\bf \nabla }\phi
\cdot \frac{\partial f_{o\alpha }}{\partial {\bf v}}
\end{equation}
in the time dependent following form[7,8]

\begin{equation}
f({\bf r},{\bf v},t)=\frac{q_{\alpha }}{m_{\alpha }}\int_{-\infty }^{t}\exp
\left[ i\omega (t-t^{\prime })\right] {\bf \nabla }\phi `(r(t^{\prime
}))\cdot \frac{\partial f_{o\alpha }}{\partial {\bf v(}t^{\prime }{\bf )}}%
dt^{\prime }
\end{equation}

\noindent where $\alpha =e,i,d.$ Now, the dispersion relation in terms of
the dielectric susceptibilities, in the low frequency approximation  ($%
\omega ,k_{z}v_{T}\ll \omega _{c}$) is

\begin{equation}
1+\sum_{\alpha }\chi _{o\alpha }=0
\end{equation}
where,

\begin{equation}
\chi _{o\alpha }=\frac{1}{(k\lambda _{D\alpha })^{2}}\left[ 1+l_{\alpha }%
\frac{\omega }{\sqrt{2}k_{z}v_{T\alpha }}{\bf Z}(\xi _{\alpha }{\bf )}%
I_{o}(z_{\alpha })e^{-z_{\alpha }}\right]
\end{equation}
with :
\begin{eqnarray}
l_{\alpha } &=&1-\frac{k_{y}T_{\alpha }}{m_{\alpha }\omega \omega _{c\alpha }%
}\left( \frac{d}{dx}\ln n_{o\alpha }+\frac{dT_{\alpha }}{dx}\frac{\partial }{%
\partial T_{\alpha }}\right)  \nonumber \\
z_{\alpha } &=&\frac{k_{y}^{2}T_{\alpha }}{m_{\alpha }\omega _{c\alpha }^{2}}
\nonumber \\
\xi _{\alpha } &=&\frac{\omega }{\sqrt{2}k_{z}v_{Te}^{2}}  \nonumber
\end{eqnarray}
Further, in order to simplify our expressions, we use:

\begin{equation}
\frac{d}{dT_{\alpha }}(\frac{1}{v_{T\alpha }})=-\frac{m_{\alpha }^{1/2}}{2\
T_{\alpha }^{3/2}};\ \ \ \frac{dz_{\alpha }}{dT_{\alpha }}=\frac{k_{y}^{2}%
}{m_{\alpha }\omega _{c\alpha }^{2}};\ \ \ \frac{d\xi _{\alpha }}{%
dT_{\alpha }}=-\frac{\omega }{2k_{z}v_{T\alpha
}^{2}\sqrt{2m_{\alpha }T_{\alpha }}}
\end{equation}
Now, using the following identity for the dispersion function ${\bf Z}$
\[
\fbox{$\displaystyle {\bf {Z^{\prime }}=-2[1+\xi _{\alpha }{Z(\xi _{\alpha })%
}],}$\nonumber}
\]
we obtain after several cumbersome algebraic manipulations the dielectric
susceptibility in the form
\begin{equation}
\chi _{o\alpha }=\frac{1}{(k\lambda _{D\alpha })}\left[ 1+\frac{\omega {\bf Z%
}I_{0\alpha }e^{-z_{\alpha }}}{\sqrt{2}\ k_{z}v_{T_{\alpha }}}\left\{ 1-%
\frac{k_{y}T_{\alpha }}{m_{\alpha }}\left( \frac{n_{0\alpha }^{\prime }}{%
n_{0\alpha }}+T_{\alpha }^{\prime }\left[ -\sqrt{\frac{m_{\alpha }}{%
T_{\alpha }^{3}}}\frac{v_{T\alpha }}{2}+\frac{{\bf Z^{\prime }}\xi ^{\prime }%
}{{\bf Z}}+\frac{I_{0}^{\prime }z_{\alpha }^{\prime }}{I_{0}}-z_{\alpha
}^{\prime }\right] \right) \right\} \right] \,
\end{equation}

In order to put our dispersion relation in a dimensionless form,
we introduce following suitable definitions:
\begin{eqnarray}
\lambda _{D\alpha } &=&\sqrt{\frac{T_{\alpha }}{n_{o\alpha }Z_{\alpha
}^{2}e^{2}}};\ \ K=k\lambda _{Di};\ \ \mu _{\alpha }=\frac{n_{o\alpha }}{%
n_{oi}}  \nonumber \\
\Theta _{\alpha } &=&\frac{T_{\alpha }}{T_{i}};\ \ \ \omega _{c\alpha }=%
\frac{Z_{\alpha }e B}{m_{\alpha }};\ \ \ k\lambda _{D\alpha }=K\sqrt{\frac{%
\Theta _{\alpha }}{\mu _{\alpha }}}  \nonumber \\
\Omega  &=&\frac{\omega }{\omega _{pi}};\ \ \ \Omega _{c\alpha }=\frac{%
\omega _{c\alpha }}{\omega _{pi}};\ \ \ U_{\alpha }=\frac{v_{T\alpha }}{%
c_{si}}  \nonumber
\end{eqnarray}
Now, using those results and assuming that $ \omega \ll
\omega_{oi}\ll \omega_{od}$ we can write down Eq.(3) as
\begin{equation}
1 + \chi_{0e} + \chi_{0i} + \chi_{0d}=0
\end{equation}
 In the non ideal case (dust)
we introduce a relation that in principle express the non ideal
behavior of the system in terms of the pressure in the form
\begin{equation}
p= n^{0}_{d} T_{d} (1 + b_{d} n^{0}_{d})
\end{equation}
given by the hard-core model. This model is taken for simplicity.
A better model, as mentioned before, will be introduced in a
future work. Now, following definitions are also useful
\begin{equation}
 \frac{1}{L_{p}}= \frac{\nabla p_{d}}{p_{d}};\ \ \frac{1}{L{n}}=
\frac{\nabla n^{0}_{d}}{n^{0}_{d}}; \frac{1}{L{d}}= \frac{\nabla
T_{d}}{T_{d}}
 \end{equation}
 Those relations are very convenient by writing the full
 dispersion relation[4]. In fact we have
\begin{equation}
\frac{1}{L_{p}}= \frac{1 + 2b_{d}n_{0d}}{1 + b_{d}n_{0d}}
\frac{1}{L_{n}} + \frac{1}{L_{T}},
\end{equation}
for the non-ideal case. For the ideal one, we use the well known
relation $ p_{0j}= n_{0j} T_{j} $, and in a similar way we get
\begin{equation}
\frac{1}{Lp_{j}}= \frac{1}{Ln_{j}} + \frac{1}{L_{T_{j}}}
\end{equation}
where $ j= i, e $.
Two special cases can be worked out:\\
A) Density gradient equal to zero $\nabla n_{0j}=0 $, that means,
$ Lp_{j} = L_{T_{j}}. $ \\
\\
B) Temperature gradient equal to zero $ \nabla T_{j}= 0 $, that
means, $ Lp_{j} = Ln_{j} . $ \\
Further we can introduce following relations in order to express
dielectric susceptibilities in a suitable forms
\begin{equation}
\frac{n'_{0j}}{n_{0j}} = \frac{1}{Ln_{j}}\equiv \frac{1}{\Lambda
n_{j} \lambda_{Di}}
\end{equation}

\begin{equation}
 T'_{j} = \frac{T_{j}}{L_{T_{j}}}\equiv
\frac{\Theta_{j} T_{i}}{\Lambda_{_{T_{j}}} \lambda_{Di}}
\end{equation}

Using those relations we arrive to the dispersion relation for the
case B where we get:
\begin{equation}
\chi_{0e}= \frac{\mu_{e}}{K^{2} \Theta_{e}}\left[1 + \frac{\Omega\
\mathbf{Z_{e}}\ I_{0e} e^{-z_{e}}}{\sqrt{2} K_{z} U_{e}} \left\{1
- \frac{K_{y} U_{e}^{2}}{\Omega\
\Omega_{0e}}\frac{1}{\Lambda_{n_{e}}} \right\} \right]
\end{equation}

\begin{equation}
\chi_{0d}= \frac{\mu_{d} Z_{d}^{2}}{K^{2} \Theta_{d}}\left[1 +
\frac{\Omega\ \mathbf{Z_{d}}\ I_{0d} e^{-z_{d}}}{\sqrt{2} K_{z}
U_{d}} \left\{1 - \frac{K_{y} U_{d}^{2}}{\Omega\
\Omega_{0d}}\frac{1}{\Lambda_{n_{d}}} \right\} \right]
\end{equation}

\begin{equation}
\chi_{0i}= \frac{1}{K^{2}}\left[1 + \frac{\Omega\ \mathbf{Z_{i}}\
I_{0i} e^{-z_{i}}}{\sqrt{2} K_{z} U_{i}} \left\{1 - \frac{K_{y}
U_{i}^{2}}{\Omega\ \Omega_{0i}}\frac{1}{\Lambda_{n_{i}}} \right\}
\right]
\end{equation}

where $ \Lambda_{n_{d}}= [(1 + 2 b_{d} n_{0d})/(1 +  b_{d}
n_{0d})]\Lambda_{p} $ and $ \Lambda_{n_{j}} = \Lambda_{p_{j}} $ .

In a similar way, it is possible to include the terms for case A,
where we shall have

\begin{equation}
\Lambda_{n_{j}} = \Lambda_{p_{j}}.
\end{equation}

Introducing now the multipolar approximation to $\mathbf{Z}$ we
can get a polynomial expression in the well known form[9]
\begin{equation}
\sum_{i}a_{i}\Omega^{i}/\sum_{j}b_{j}\Omega^{j} = 0
\end{equation}
where coefficients $a_{i}$ and $b_{i}$  are functions of the
system parameters. Such an expression is easy to solve and with
high accuracy to find roots of the numerator. An analysis of these
solutions spectra permit us to give the imaginary parts $ \gamma =
Im(\Omega)$ in function of $ 1/K_{y}$, which represent the growth
rate instabilities.

\section{Results and Conclusions}

The quasi-neutrality equation for dusty plasmas can be approached by a
simplified one due to the high state of charge of the dust grains
\begin{equation}
n_{oi}=Z_{D}n_{od}+n_{oe}\simeq Z_{D}n_{od}
\end{equation}
and the electron susceptibility can be neglected in the dispersion relation.
The range of the main parameters in the study of the low frequency
oscillation of dust grains is established by the approximations that
conduced to the simplified dispersion relation
\begin{equation}
\Omega ,K_{z}U_{d}\ll \Omega _{cd}
\end{equation}
Unstable dust oscillations (${Im}(\Omega )>0$) are found for $\Omega
_{cd}\simeq 10^{-1}$, $K_{z}U_{d}\simeq 10^{-2}$. At the present time, we
only give the results for the density gradient case ({\em i.e.} $\partial
/\partial T=0$). For slightly inhomogeneous plasmas with normalized density
gradient length $\Lambda _{n}=n_{o\alpha }/(\lambda _{Di}\nabla n_{o\alpha
})\approx 10^{2}$, the shape of the dust instability (${Im}(\Omega )_{\max }$%
) curve as function of the perpendicular to magnetic field wavelength ($%
1/K_{y}$) is similar to that for ions, previously studied [8].

\begin{figure}
\begin{center}
\includegraphics[height=9.5cm, width=9.5cm]{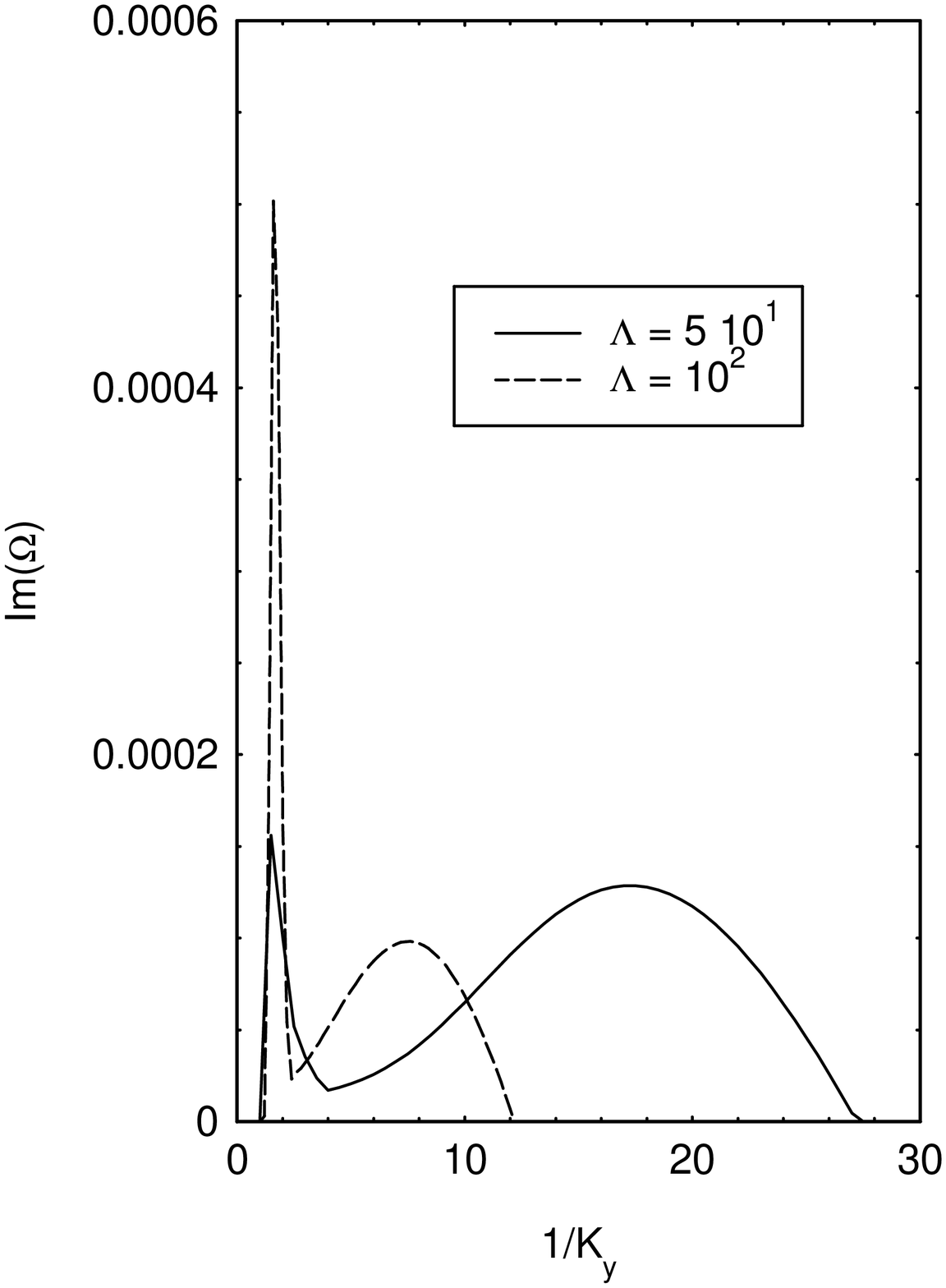}
\caption{Normalized maximum growth rate as a function of
normalized perpendicular wavelength for slightly inhomogeneous plasma ($%
\Lambda = 10^{2}$) and for a relatively inhomogeneous one ($\Lambda =
5\times 10^{1}$).}
\end{center}
\end{figure}

The maximum value of the instability increases and narrows with
the state of charge of the dust $Z_{D}$ but decreases and get
wider with the mass. For typical laboratory light dusty plasmas
($m_{d}\sim 10^{4}m_{p}$, $Z_{D}\sim 10^{3}$) the instability of
dust acoustic or electrostatic waves is narrower and smaller than
that for ions In figure 1 the peak of the left corresponds to the
typical shape of instability of slightly inhomogeneous plasmas,
while the right region of instability appears for density gradient
lengths of the order of a hundred of Debye lengths ($\Lambda
_{n}\equiv \Lambda \lesssim 10^{2}$). For higher density gradients
($\Lambda =5\times 10^{1}$), this new instability region is wider
and so high as the typical one. For even higher density gradients
($\Lambda =10^{1}$), figure 2 shows that the new right region
gives a higher instability. This figure also shows the effect of
the non ideality of the plasma. necessary condition for the
exhibition of dust acoustic waves. For typical laboratory dust
radius the $b_{od}$
parameter of the hard core potential equation of state, is of the order of $%
10^{-14}m^{3}$. And for typical values of ion density of
$10^{16}m^{-3}$ (and corresponding $n_{od}$, by quasi-neutrality
relation), it appears a new intermediate instability region which
can reach a maximum for denser plasmas ($n_{oi}=10^{17}m^{-3}$) or
larger dust particles ($b_{od}\gtrsim 10^{-13}m^{3}$). This
maximum is limited for the relation for dust collective behavior
\begin{equation}
r_{d}\ll \lambda _{Di}
\end{equation}

\begin{figure}
\begin{center}
\includegraphics[height=9cm, width=9.5cm]{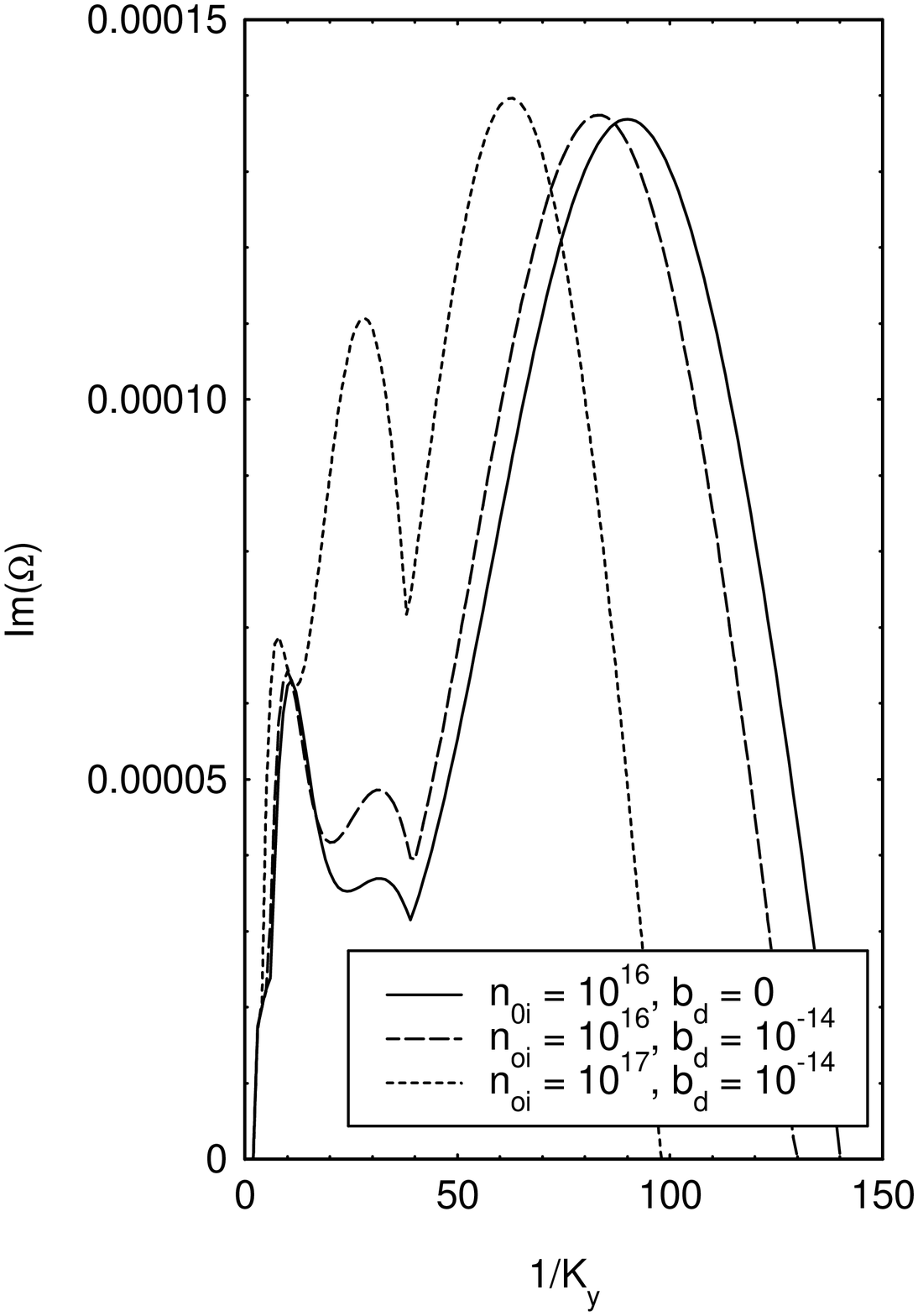}
\caption{Normalized maximum growth rate as a function of
normalized perpendicular wavelength for ideal and non ideal plasmas.}
\end{center}
\end{figure}

\section{References}

\begin{enumerate}
\item J. H. Chen, J. B. Du, and Lin I., {\it J. Phys}. D: Appl.
Phys. \textbf{27}, 296(1994)

\item A. Barkan, R. L. Merlino, and N. D'Angelo, {\it Phys.
Plasmas}, \textbf{2}, 3563(1995)

\item Reichl L.E., {\ {\it ''A Modern Course in Statistical
Physics''}}, Edward Arnold, 1991.

\item N. N. Rao, "{\it Frontier in Dusty Plasmas}", Y. Nakamura,
T. Yakota, and P.K. Shukla, Eds., Elsevier Science B.V,(2000)

\item F. H. Ree and W. G. Hoover, {\it J. Chem. Phys},
\textbf{40}, 939(1964)

\item  P. Mart\'{i}n et al., J. Math. Phys. {\bf 21}, 280 (1980)

\item  Mikhailovsky A. B., {\ {\it ''Handbook of Plasma
Physics''}}, Rosenbluth and Sagdeev Eds., North Holland,
Amsterdam, 1983.

\item  A. Galeev et al. Soviet Physics. JETP {\bf 17}, 615 (1963)

\item  J. Puerta and C. Cereceda, Proc. ICPPP {\bf 1}, 94 - 97
(1996)

\end{enumerate}
\end{document}